\documentclass{appolb}
\usepackage{epsfig}

\begin{document}
\title{Detection of Supernova Neutrinos%
\thanks{Presented at the XXVIII Mazurian Lakes School of Physics, Krzy¿e, Poland,\break
August 31--September 7, 2003.}
\thanks{This work has been partly supported by the Polish State Committee for Scientific
        Research (KBN) under Contracts No. 2 P03B10925 and
160/E-340/SPB/ICARUS/\break P-03/DZ212/2003-2005.}%
}
\headtitle{Detection of Supernova Neutrinos}
\author{B. Bekman, 
J. Holeczek and 
J. Kisiel
\address{Institute of Physics, University of Silesia\\ 
Uniwersytecka 4, 40-007 Katowice, Poland}
}
\headauthor{B.~Bekman, J.~Holeczek, J.~Kisiel}
\maketitle

\begin{abstract}
Matter effects on neutrino oscillations in both, a supernova and the Earth, change the
observed supernova neutrino spectra.
We calculate the expected  number of supernova neutrino interactions for ICARUS, SK and
SNO detectors
as a function of the distance which they traveled in the Earth. Calculations are performed
for
supernova type II at 10~kpc from the Earth, using standard supernova neutrino fluxes
described
by thermal Fermi--Dirac distributions and the PREM I Earth matter density profile.
\end{abstract}


\section{Introduction}

Iron is the most strongly bound of all elements
which means that fusion and fission reactions result in the absorption,
rather than the production, of energy.
The last stage of a massive star life (mass $>10M_{\odot }$) is the
collapse of its Fe core (the whole core material has already been
transformed, via the chain of nuclear reactions, into Fe).
This happens when the mass of the Fe core exceeds the Chandrasekhar limit
($\sim 1.45M_{\odot }$ if one takes equal numbers of neutrons and protons, \cite{bethe})
--- a supernova (SN) type II birth is a fact. It starts to explode. All flavors
of neutrinos are radiated away in the form of two bursts of the duration
of milliseconds and seconds. The emitted neutrinos carry almost all
($\sim $99\%) of the SN binding energy ($\sim $10$^{53}$~erg).

The only observed, up to now, burst of SN neutrinos came from the
SN1987A which had exploded in the Large Magellanic Cloud, at the distance
of about 52~kpc away from the Earth. Due to this distinct distance, the
reconstruction of only 19 events of the neutrino interactions by the
Kamiokande \cite{kamiokande} and IMB \cite{imb} water
Cerenkov detectors had been possible. But, it was enough to confirm
the main features of the models of SN explosion.

Neutrinos, on their way from the production point inside the SN
high dense Fe core to the terrestrial detector, interact with matter.
Non-zero neutrino masses, together with flavor mixing enhanced by
matter effects, result in considerable differences in neutrino
fluxes between the production and the detection points.
Systematic studies of the number
of SN neutrino interactions in three detectors, ICARUS \cite{ICARUS}, SK \cite{SK} and
SNO \cite{SNO}, as a function of the distance passed by neutrinos in the Earth is the
aim of this work.

\section{Neutrino fluxes from a supernova}

A supernova is a source of fluxes of neutrinos and antineutrinos of all three
flavors ($e$, $\mu$, $\tau$).
The neutrino/antineutrino energy spectra at the production point (\ie
inside the SN) of various flavors can be described by the thermal Fermi--Dirac
distributions
(all chemical potentials set to zeros, see \cite{smirnov}):
\begin{eqnarray*}
F^0_\alpha(E,T_\alpha,L_\alpha)=
\frac{ L_\alpha}{ T^4_\alpha F_3}\,
\frac{ E^2}{ \e^{E\slash T_\alpha} + 1}\,,
\end{eqnarray*}
where $\alpha =\nu_{e}$, $\nu_{\mu }$, $\nu_{\tau }$, $\bar \nu_{e}$, $\bar \nu_{\mu}$,
$\bar \nu_{\tau}$.
Here, $E$ represents energy of the neutrinos,
$L_\alpha$ is the total energy released in various flavors of neutrinos ($L_\alpha \simeq
E_{\rm B}/6$, with $E_{\rm B} \simeq 3 \times 10^{53}$~ergs --- binding energy emitted in the core
collapse of the star),
$F_3$ is a normalizing constant given by $F_3=7\pi^4/120$, $T_\alpha$ is the temperature
of the $\nu_\alpha$ gas in the neutrino sphere. We assume the following hierarchy of
temperatures: $T_{\nu_e} = 3.5$~MeV, $T_{\bar {\nu}_{e}} = 5$~MeV, $T_{\nu_x,\bar{\nu}_x}
= 8$~MeV, where $\nu_x$ and  $\bar{\nu}_x$ mean $\nu_{\mu}$, $\nu_{\tau}$ and
$\bar{\nu}_{\mu}$, $\bar{\nu}_{\tau}$, respectively.

Neutrinos produced deep inside supernova, before escaping from the star, interact with the
matter of its mantle.
Thus, transitions of neutrino species can occur.
These oscillations of neutrinos in supernova are considered here according to \cite
{sn_neutrino_burst}.
The conversions take mainly place in two resonance layers in the outer
regions of a supernova mantle.
There is a high density resonance region (H resonance layer) --- 
$\rho_{\rm H}\approx 10^3$--$10^4~{\rm g/cm}^3$,
and low density resonance region (L resonance layer) --- 
$\rho_{\rm L}\approx 10$--30~g/cm$^{3}$.
Transitions in the H resonance layer are governed by atmospheric neutrino oscillation
parameters: $\Delta m^2_{31}$ and ${\it\Theta}_{13}$, whereas transitions in the L resonance
region are governed by solar neutrino oscillation parameters: $\Delta m^2_{21}$ and
${\it\Theta}_{12}$.
The transition probabilities (between two neutrino mass eigenstates) in the resonance
layers are called the {\it {flip probabilities}} --- $P_{\rm H}$, $P_{\rm L}$ for neutrinos and
$\bar{P}_{\rm H}$, $\bar{P}_{\rm L}$ for antineutrinos.

The final total neutrino/antineutrino mass eigenstates fluxes on a supernova surface
($F_i, F_{\bar{i}},~i = 1,2,3$) are given by the following sets of equations
($F^0_\alpha$, $F^0_{\bar \alpha}$ are the initial total production fluxes defined above,
respectively, $e$ --- $\nu_{e}$, $\bar{e}$ --- $\bar \nu_{e}$, $x$ --- $\nu_{\mu}$ or
$\nu_{\tau}$, $\bar{x}$ --- $\bar{\nu}_{\mu}$ or $\bar{\nu}_{\tau}$).

\noindent For Direct (normal) mass hierarchy ($m_1 < m_2 \ll m_3$):

\begin{eqnarray*}
F_1 & = & P_{\rm H} P_{\rm L} F^0_e + (1-P_{\rm H} P_{\rm L})
F^0_x\,,\\[2mm]
F_2 & = & (P_{\rm H} - P_{\rm H} P_{\rm L}) F^0_e 
+ (1 - P_{\rm H} + P_{\rm H} P_{\rm L}) F^0_x\,,\\[2mm]
F_3 & = & (1 - P_{\rm H}) F^0_e + P_{\rm H} F^0_x\,,\\[2mm]
F_{\bar{1}} & = & (1 - \bar{P}_{\rm L}) F^0_{\bar{e}} 
+ \bar{P}_{\rm L} F^0_{\bar{x}}\,,\\[2mm]
F_{\bar{2}} & = & \bar{P}_{\rm L} F^0_{\bar{e}} 
+ (1 - \bar{P}_{\rm L}) F^0_{\bar{x}}\,,\\[2mm]
F_{\bar{3}} & = & F^0_{\bar{x}}\,.
\end{eqnarray*}
For Inverted mass hierarchy ($m_3 \ll m_1 < m_2$):

\begin{eqnarray*}
F_1  & = &  P_{\rm L} F^0_e + (1 - P_{\rm L}) F^0_x\,,\\[2mm]
F_2  & = &  (1 - P_{\rm L}) F^0_e + P_{\rm L} F^0_x\,,\\[2mm]
F_3   & = &  F^0_x\,,\\[2mm]
F_{\bar{1}}   & = &   (\bar{P}_{\rm H} - \bar{P}_{\rm H} \bar{P}_{\rm L}) F^0_{\bar{e}} + (1 - \bar{P}_{\rm H} +
\bar{P}_{\rm H} \bar{P_{\rm L}}) F^0_{\bar{x}}\,,\\[2mm]
F_{\bar{2}}   & = &   \bar{P}_{\rm H} \bar{P}_{\rm L} F^0_{\bar{e}} + (1 - \bar{P}_{\rm H} \bar{P}_{\rm L})
F^0_{\bar{x}}\,,\\[2mm]
F_{\bar{3}}   & = &   (1 - \bar{P}_{\rm H}) F^0_{\bar{e}}
 + \bar{P}_{\rm H} F^0_{\bar{x}}\,.
\end{eqnarray*}

A complete discussion of how to calculate flip probabilities can be found in \cite
{sn_neutrino_burst}.
For the purpose of this paper we will only state here that, in case one considers the
LMA (Large Mixing Angle) neutrino oscillations parameters: $P_{\rm L} = \bar{P}_{\rm L} = 0$ and,
in the so called $Large \enspace {\it\Theta}_{13}$ case
($\sin^2{{\it\Theta}_{13}} >  3 \times 10^{-4}$, the so called region~I in \cite
{sn_neutrino_burst}): $P_{\rm H} = \bar{P}_{\rm H} = 0$,
while in the so called $Small \enspace {\it\Theta}_{13}$ case
($\sin^2{{\it\Theta}_{13}} <  2 \times 10^{-6}$, the so called region~III in \cite
{sn_neutrino_burst}): $P_{\rm H} = \bar{P}_{\rm H} = 1$ (it is interesting to notice here that, in
this case, the resulting fluxes are equal for both, Direct and Inverted, mass
hierarchies).
%($\sin^2{\Theta_{13}} <  2 \times 10^{-6}$, the so called region III in 
%\cite {sn_neutrino_burst}): $P_H = \bar{P}_H = 1$ (it is interesting to notice here 
%that, in this case, the resulting fluxes are equal for both, Direct and Inverted, 
%mass hierarchies).

The finite spread of the neutrino wave packets, together with the small value of their
coherence length and the large distance from supernova to the Earth, imply that neutrinos
arrive to the surface of the Earth as fluxes of incoherent mass eigenstates.

\newpage
Thus, the above equations describe also (except for a simple geometrical factor related to
the distance between a supernova and the Earth) the neutrino/antineutrino mass
eigenstates fluxes at the surface of the Earth (because there is no matter on the way
between a supernova and the Earth, there can be no additional transitions between
neutrino mass eigenstates).

\section{Earth matter effect}
In order to calculate oscillation probabilities we used the standard description of
neutrino regeneration effect in the Earth \cite{MatterEffects}.
The numerical calculations were made with use of the CERN library function DEQBS.
We applied the realistic Earth matter density profile PREM I \cite{PREMI,dziewonski}.

The following LMA I (Large Mixing Angle) neutrino oscillation parameters were used
(note the two values of the $\sin^2{{\it\Theta}_{13}}$):
\begin{eqnarray*}
\Delta m^2_{21} & = & 7.1 \times 10^{-5}~{\rm eV}^2\,,\\
\Delta m^2_{32} & = & 2.5 \times 10^{-3}~{\rm  eV}^2\,,\\
\sin^2{2{\it\Theta}_{12}}& = &0.84\,,\\
\sin^2{2{\it\Theta}_{23}}& = &1.0\,,\\
\sin^2{{\it\Theta}_{13}}& = &0.02 \hspace{2.0cm} Large \enspace {\it\Theta}_{13}\,,\\
\sin^2{{\it\Theta}_{13}}& = &10^{-7} \hspace{1.9cm} Small \enspace {\it\Theta}_{13}\,,\\
\delta & = & 0 \hspace{2.5cm} Dirac's \enspace phase\,.
\end{eqnarray*}

In addition we also considered two cases of mass hierarchies,
the Direct (normal, $m_1 < m_2 \ll m_3$) and the Inverted ($m_3 \ll m_1 < m_2$) ones
(however, because the energies of SN neutrinos are relatively small and the Dirac's phase
$\delta$ is set to zero, there are almost no differences in results of the regeneration in
the Earth between these two hierarchies).

\section{Detection of supernova neutrinos}
%rys.1
\begin{figure}[htb]
\centerline{%
\epsfig{file=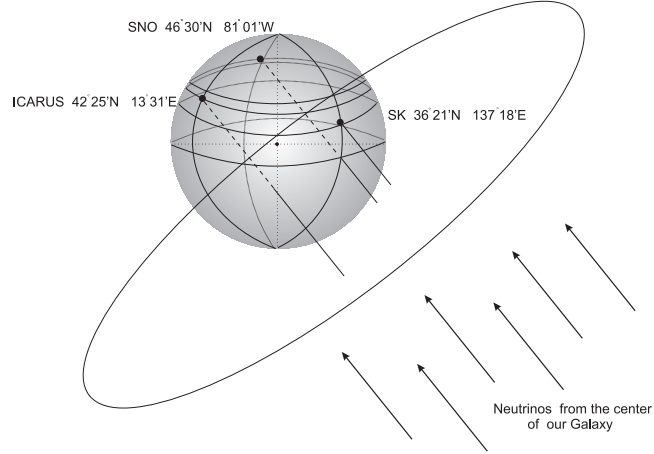,width=8.5cm}}
\caption{An example of a possible relative position of ICARUS, SK and
SNO detectors and a supernova which exploded in the center of our Galaxy.}
\label{fig.crossing_earth}

\end{figure}

We consider three detectors: ICARUS \cite{ICARUS}, SK \cite{SK} and SNO \cite{SNO}.
The positions of these detectors on the Earth are shown in Fig.~\ref{fig.crossing_earth}.
These detectors have the following feature: in most of the time, when a possible neutrino
signal arrives to the Earth, at least one detector is shielded by the
Earth (assuming a supernova exploded in the center of our Galaxy), and therefore we will see the
regeneration effect in the Earth.
All possible processes which contribute to the total number of neutrino interaction in
each detector are taken into account. The cross sections for these processes which we use
come from \cite{cross} and \cite{cross1}.

\section{Results and discussion} 

\begin{figure}[htb]
\centerline{\epsfig{file=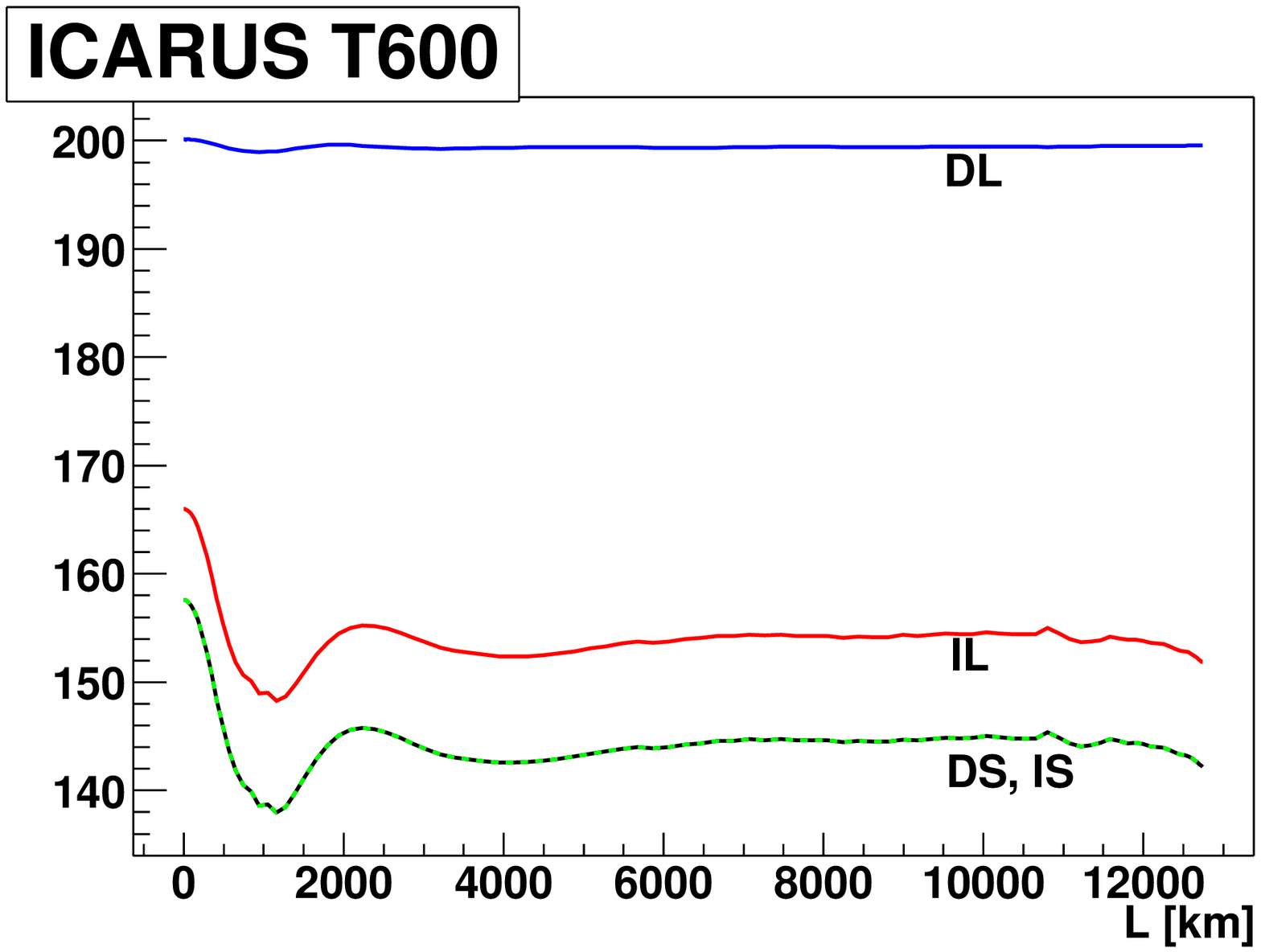,height=4.8cm}}

\centerline{\epsfig{file=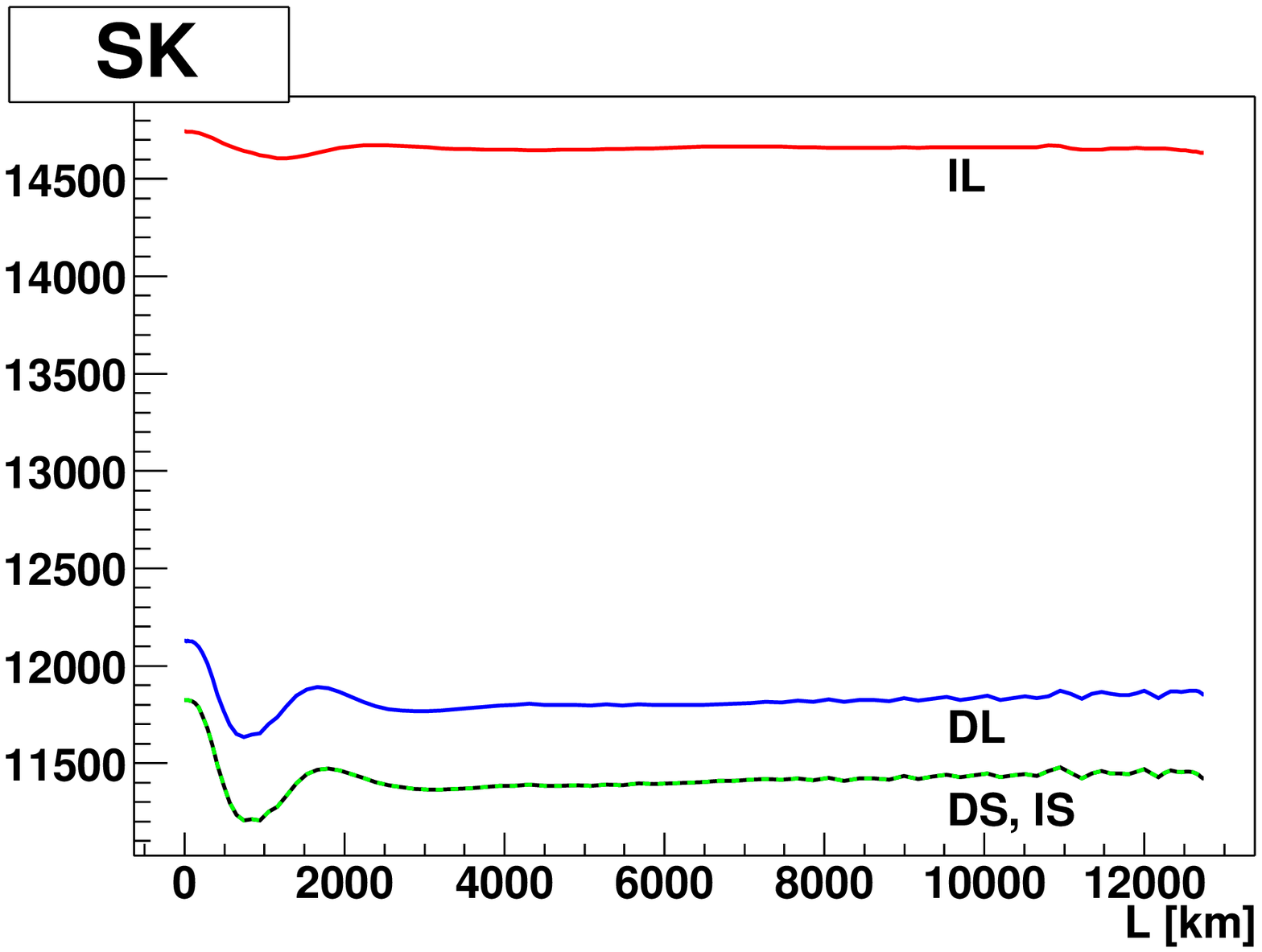,height=4.8cm}}

\centerline{\epsfig{file=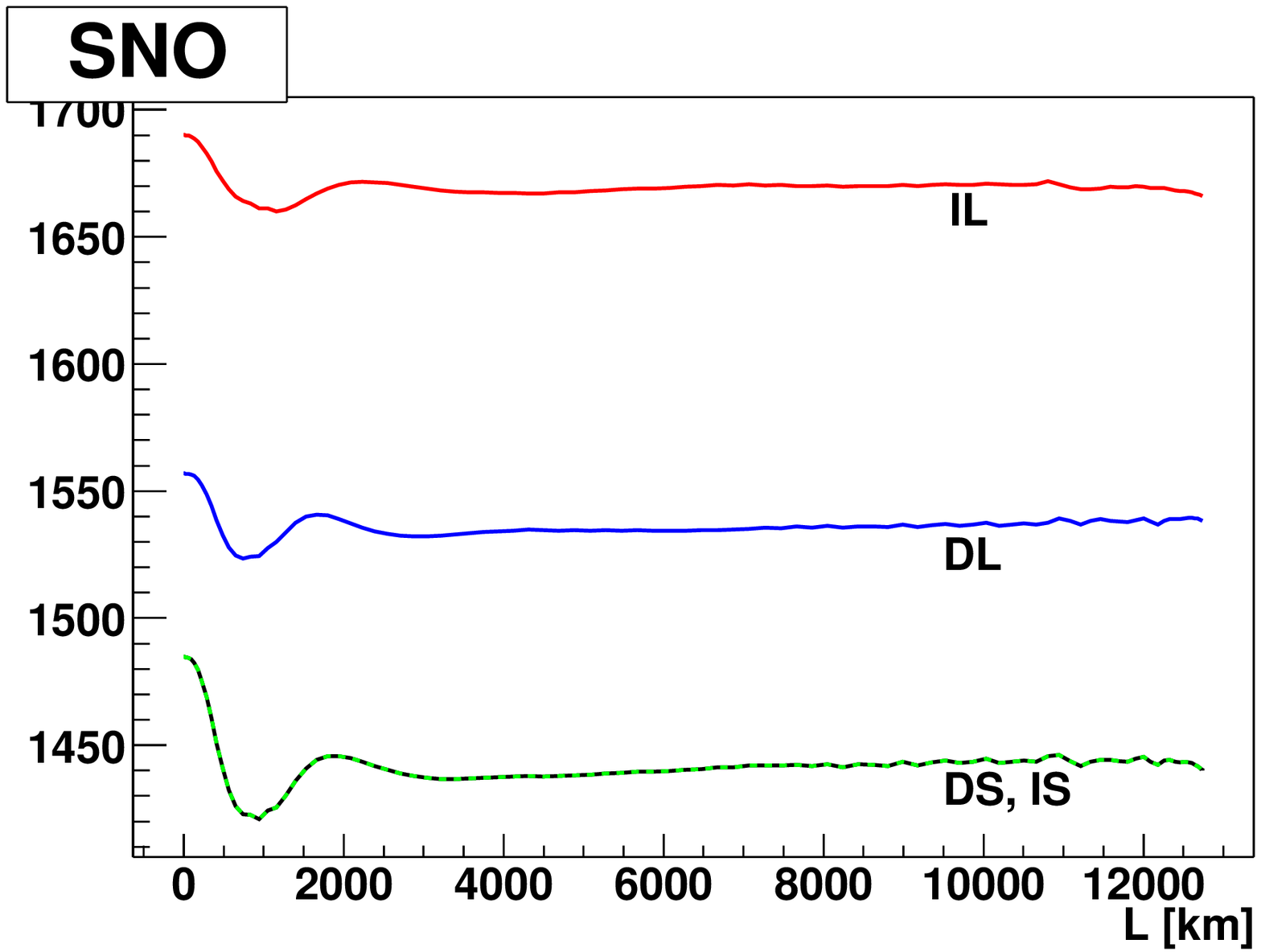,height=4.8cm}}

\caption{The expected number of supernova neutrino interactions
in ICARUS T600 (600 tons of liquid argon), SK (32~ktons of light water)
and SNO (1~kton of heavy water, 1.7~ktons of light water) detectors as a
function of the distance, L, neutrinos traveled in the Earth, for
different combinations of mass hierarchy and ${\it\Theta}_{13}$ (DL --- Direct
mass hierarchy and $Large \enspace{\it\Theta}_{13}$, IL --- Inverted mass
hierarchy and $Large \enspace{\it\Theta}_{13}$, DS --- Direct mass hierarchy
and $Small \enspace{\it\Theta}_{13}$, IS --- Inverted mass hierarchy and
$Small \enspace{\it\Theta}_{13}$). For details see the description in the
text.}

\label{fig.ICARU_SK_SNO}
\end{figure}

The expected total numbers of
supernova neutrino interactions $N_{\rm SN}$, integrated over neutrino
energy in the range 0.1~MeV--100~MeV (the whole supernova neutrino energy
spectrum), for all possible neutrino processes in ICARUS T600 (an
``industrial'' ICARUS module filled with 600 tons of liquid argon), SK
(32~ktons of light water) and SNO (1~kton of heavy water, 1.7~ktons of
light water) detectors are calculated (neutrino oscillations in
supernova and the regeneration effect in the Earth have been taken into
account). It is assumed that a supernova explosion occurred in the
center of our Galaxy, that is 10kpc away from the Earth. The results are
presented in Fig.~\ref{fig.ICARU_SK_SNO} as a function of the distance
which neutrinos traveled in the Earth, for four possible combinations of
the mass hierarchy and the ${\it\Theta}_{13}$ value.

%rys.2

The main neutrino interactions with detector materials which contribute to the $N_{\rm SN}$
are the following
(the minimum and maximum contributions of a particular process, taken for the four
considered cases altogether, into the total number of interactions are also given):
\begin{table}[htb]

\centerline{\small
\begin{tabular}{lrcll}
ICARUS 
&$\nu _{e}+^{40}$Ar&$\rightarrow$ &${}^{40}{\rm K}^{*}+e^{-}$ &$\sim (87 \div 93)$\% \\
[2mm]
SK 
&$\bar \nu _{e}+p$&$\rightarrow$ &$n+e^+$ &$\sim (76 \div 80)$\%\\
&$\nu _{e}+{\rm O}$&$\rightarrow$ &${\rm F}+e^-$ &$\sim (6 \div 10)$\%\\
[2mm]
SNO 
&$\bar \nu _{e}+p$&$\rightarrow$ &$n+e^+$ &$\sim (31 \div 37)$\%\\
&$\nu _{e}+d$&$\rightarrow$ &$p+p+e^-$ &$\sim (12 \div 16)$\%\\
&$\nu _{\mu,\tau}+d$&$\rightarrow$ &$\nu _{\mu,\tau}+p+n$ &$\sim (10 \div 12)$\%\\
&$\bar \nu _{\mu,\tau}+d$&$\rightarrow$ &$\bar \nu _{\mu,\tau}+p+n$ &$\sim (8 \div
12)$\%
\end{tabular}
}
\end{table}

It can be seen that, while $\nu_{e}$ interactions dominate in the ICARUS detector,
$\bar \nu_{e}$ interactions dominate in the SK and SNO detectors.
Taking into account that, in case of $Large \enspace{\it\Theta}_{13}$,
neutrino oscillations in supernova make the $\nu_{e}$ spectrum harder (hot) for the 
Direct mass hierarchy
and the $\bar \nu_{e}$ spectrum harder (hot) for the Inverted mass hierarchy
(and that all relevant cross sections increase with energy), one gets
the corresponding behavior of $N_{\rm SN}$ in Fig.~\ref{fig.ICARU_SK_SNO} 
(compare the DL {\it versus} the IL curves).
%(the highest values of $N_{SN}$ for the Direct mass hierarchy and the ICARUS detector, and for the Inverted mass hierarchy and the SK and SNO detectors).

Two conclusions are straightforward: the distance traveled by neutrinos
in the Earth has only little influence on the value of $N_{\rm SN}$ and, in
case of $Small \enspace{\it \Theta}_{13}$, the value of $N_{\rm SN}$ does not
depend on the mass hierarchy at all. Finally, the $N_{\rm SN}$ from all
three detectors should allow us to draw conclusions about the value of
the ${\it\Theta}_{13}$ and, in case the ${\it\Theta}_{13}$ is sufficiently large,
it should also be possible to say which mass hierarchy is in force.

Last, but not least, it should be noted that, in case of the ICARUS detector, for the
purpose of this paper we performed calculations for the currently existing ICARUS T600
(600 tons of liquid argon) module. The final total mass of the ICARUS detector (which will
be installed in the underground LNGS Laboratory in Gran Sasso/Italy, see \cite{ICARUS})
will be of the order of 3000 tons of liquid argon. That means that the expected total
number of supernova neutrino interactions $N_{\rm SN}$ will be five times larger than the one
presented in this paper.

\end{document}